\documentclass[10pt,aps,amsmath,showpacs,amssymb,nofootinbib,pre,twocolumn]{revtex4}
\usepackage[dvips]{graphicx}
\usepackage[dvips,usenames]{color}
\usepackage{amsmath}

\newcommand{\av}[1]{\langle {#1} \rangle}

\begin{document}

\title{Spectral analysis and slow spreading dynamics on complex networks} 

\author{G\'eza \'Odor}
\affiliation{Research Centre for Natural Sciences, 
Hungarian Academy of Sciences, MTA TTK MFA, 
P. O. Box 49, H-1525 Budapest, Hungary}

\pacs{89.75.Hc, 05.70.Ln, 89.75.Fb}
\date{\today}

\begin{abstract}

The Susceptible-Infected-Susceptible (SIS) model is one of the simplest 
memoryless system for describing information/epidemic spreading phenomena 
with competing creation and spontaneous annihilation reactions.
The effect of quenched disorder on the dynamical behavior has recently been 
compared to quenched mean-field (QMF) approximations in scale-free networks.
QMF can take into account topological heterogeneity and clustering 
effects of the activity in the steady state by spectral decomposition 
analysis of the adjacency matrix. Therefore, it can provide predictions 
on possible rare-region effects, thus on the occurrence of slow dynamics.
I compare QMF results of SIS with simulations on various large dimensional
graphs. In particular, I show that for Erd\H os-R\'enyi graphs this method
predicts correctly the epidemic threshold and the rare-region effects. 
Griffiths Phases emerge if the graph is fragmented or if we apply strong, 
exponentially suppressing weighting scheme on the edges. The latter model
describes the connection time distributions in the face-to-face experiments. 
In case of generalized Barab\'asi-Albert type of networks with aging 
connections strong rare-region effects and numerical evidence for 
Griffiths Phase dynamics are shown.
\end{abstract}

\maketitle

\section{Introduction}

Dynamical processes evolving on complex networks are of current interest of 
research \cite{dorogovtsev07:_critic_phenom,barratbook}. 
In networks with large topological 
dimension defined as $N\propto r^d$, where $N$ is the number of nodes 
within the (chemical) distance $r$, the dynamics is expected to be 
exponentially fast. 
However, there are observations showing the appearance of generically 
slow time evolution. For example in working memory of the brain \cite{Johnson} 
or in recovery processes following a virus pandemic \cite{pv01a,pv04,vir} 
power-law type of time dependencies have been found, resembling of dynamical 
critical phenomena \cite{Chialvo}. In social networks the occurrence of generic 
slow dynamics was suggested to be the result of bursty behavior of agents 
connected by small world networks \cite{KK11}.

Another possible explanation is related to the emergence of arbitrarily large, 
rare-regions (RR) that can change their state exponentially slowly as the
function of their sizes. 
Near phase transitions from active to inactive states in disordered system
\cite{DickMar,Henkel,rmp,odorbook} a so called Griffiths Phase (GP) \cite{Griffiths,Vojta} 
can develop, characterized by non-universal, power-law dynamics. 
Griffiths singularities affect the dynamical behavior both below and above 
the transition point and can be best described via renormalization group methods
in networks \cite{KI11,JK13}.
It has been been conjectured \cite{PhysRevLett.105.128701,odor:172,Juhasz:2011fk}
that such slow dynamics can occur only in finite dimensional networks as the 
consequence of heterogeneity: explicit reaction or purely topological disorder.
This is based on optimal fluctuation theory and simulations of the Contact Process
(CP) \cite{harris74,liggett1985ips} on Erd\H os-R\'enyi (ER) \cite{ER} 
and on Generalized Small World networks \cite{an,Juhasz3,Juhasz}.
In case of networks with infinite topological dimension, like the 
Barab\'asi-Albert (BA)\cite{Barabasi:1999} graph slow dynamics has
be found only in tree networks and weighting schemes, that 
suppress the information propagation among hubs \cite{BAGPcikk,wbacikk}.

The Susceptible-Infected-Susceptible (SIS) model \cite{SIS} is another 
fundamental system to describe simple epidemic (information) possessing 
binary site variables: infected/active or healthy/inactive. 
Infected sites propagate the epidemic (or active) all of their neighbors 
with rate $\lambda$ or recover (spontaneously deactivate) with rate $1$. 
SIS differs from the CP in which the branching rate is normalized by $k$, 
the number of outgoing edges of a vertex, thus it allows an analytic 
treatment, using symmetric matrices.
By decreasing the infection rate of the neighbors a continuous phase transition 
occurs at some $\lambda_c$ critical point from a state with finite 
activity density $\rho$ to and inactive steady state with $\rho=0$.
The latter is also called absorbing, because no spontaneous activation of sites
is allowed. In \cite{wbacikk} it was shown that a dissortative weighing 
scheme of the edges $w_{ij}\in (0,1)$ can effectively slow down the 
information propagation in SIS model and result in long living 
rare regions, causing slow dynamics.

Heterogeneous Mean Field (HMF) theory represents an exact result in 
annealed networks and provides a good approximation in networks with 
high $d$, when the dynamical fluctuations are irrelevant 
\cite{boguna09:_langev,FCP12}. To describe quenched heterogeneity 
of the network the so-called Quenched Mean-Field (QMF) approximation
is introduced \cite{CWW08,Mieg09,GDOM12}. 
In \cite{wbacikk} I compared density decay simulation results of the
SIS model with the QMF approximations.
Here I show further evidences for the agreement 
of QMF and dynamical simulations in case of certain ER and BA graphs. 
In particular, I investigate the prediction of QMF for the interaction 
weight scheme of \cite{b13} applied for ER graphs. 
This kind of disorder on interactions is important, because face-to-face 
experiments \cite{F2F} resulted in such distribution of connect intensities
that can be modeled with it.
In \cite{BAGPcikk,wbacikk} $k$ dependent weights were applied
on the edges, while in \cite{PhysRevLett.105.128701,odor:172,Juhasz:2011fk}
the infection probability of nodes were reduced to slow down the 
fast epidemic spreading in small world networks. 
Now I apply QMF for $k$ independent weights, distributed and frozen on
the graph edges before the start of the epidemic process.

In the original BA graph construction one starts from a single connected
node and add new links with the linear preferential rule. This causes
initial nodes with high connectivity and those are attached at step $i$
will have a vanishing average degree $\langle k_i\rangle \propto 1 / i^{1/2}$. 
In various network studies, like article citations \cite{Eom} and model calculations
\cite{Dor&Men00} the degradation of connection capability of aging nodes 
have been analyzed. It is well known that in neural networks this happens indeed.
In this paper I investigate a generalized model, in which fraction of
edges of the aging nodes are removed by a random, linear preferential rule \cite{SAH11}.
In this case the edge distribution of the BA graph $P(k)\propto k^{-3}$ 
will be cut off by an exponential factor for large $k$-s and the QMF suggests 
GP behavior in agreement with the dynamical simulations.

\section{Spectral analysis and quenched mean-field approximations}

A mean-field theory of the SIS model \cite{Mieg09,GDOM12}, 
capable of taking into account the topological heterogeneity 
in a network of size $N$ is based on the rate equation of $\rho_i(t)$, 
the infection probability of node $i$ at time $t$: 
\begin{equation}
\label{qmfsis}
\frac{d\rho_i(t)}{dt} = -\rho_i(t) + \lambda (1-\rho_i(t))\sum_{j=1}^N A_{ij}w_{ij} \rho_j(t)~.
\end{equation}
Here $A_{ij}$ is an element of the adjacency matrix assigned with $1$, 
if there is an edge between nodes $i$ and $j$ or $0$ otherwise and $w_{ij}$
describes the possibility of weights attributed to the edges.
For large times the SIS model evolves into a steady state, with and order 
parameter $\rho \equiv \av{\rho_{i}}$.
This equation with symmetric weights under the exchange of 
$i \leftrightarrow j$ can be treated by a spectral decomposition (SD) 
on an orthonormal eigenvector basis. Furthermore the non-negativity of 
the $B_{ij} = A_{ij}w_{ij}$ matrix involves a unique, real, non-negative largest 
eigenvalue $\Lambda_{1}$. In the QMF approximation one can find $\lambda_c$ 
and $\rho(\lambda)$ around it by taking into account the principal 
eigenvector only. Using the linear superposition expansion of $\rho$ 
one can solve Eq.~(\ref{qmfsis}), which provides  $\Lambda_{1}=1/\lambda_c$, 
i.e. a stable $\rho >0$ (active) solution for $\lambda > \lambda_c$ and 
an inactive one for  $\lambda \le \lambda_c$.
The order parameter near, above $\lambda_c$ can be approximated via
\begin{equation}
\rho(\lambda) \approx a_1 \Delta + a_2 \Delta^2 + ... \ ,
\end{equation}
where $\Delta = \lambda \Lambda_{1}{-}1 {\ll} 1$ and the coefficients
\begin{equation}
a_j = \sum_{i=1}^N f_i(\Lambda_j)/[N \sum_{i=1}^N f_i^3 (\Lambda_j)] \,
\label{epsilon}
\end{equation}
are functions of eigenvectors of the largest eigenvalues. This expression
is exact if there is a gap between $\Lambda_1$ and $\Lambda_2$ \cite{Mieg12}. 

It was proposed in \cite{GDOM12} and tested on weighted BA models 
\cite{wbacikk} that the localization of activity in the active steady 
state can be characterized by the Inverse Participation Ratio ($IPR$),
related to the eigenvector of the largest eigenvalue 
$\mbox{\boldmath$f$}(\Lambda_1)$ as
\begin{equation} \label{IPR}
IPR(N) \equiv \sum_{i=1}^{N} f_{i}^{4}(\Lambda) 
\end{equation}
This quantity remains small: $\lim_{N\to\infty} IPR(N) = 0$ in case of 
homogeneous eigenvector components and takes the maximal value $1$ if all 
activity is concentrated on a single node. I used the sparse matrix
package OCTAVE \cite{OCT} for generating and diagonalizing $B_{ij}$ and
calculating $IPR(N)$, $\Lambda_1(N)$, $a_i(N)$ for network sizes up to $N=200.000$.
In the numerical analysis I extrapolated and fitted the $IPR(N)$ 
data assuming a power-law form
\begin{equation} \label{IPRscal}
IPR = IPR(N) + X (1/N)^{b} \ ,
\end{equation}
containing the free parameters: $X$ and $b$.

It was derived by \cite{Chung}, that the largest eigenvalue of $A_{ij}$ of 
general random graphs is determined by the maximum degree $k_{max}$ 
\begin{equation}
\Lambda_1(N) = \big(1 + o(1) \big) {\rm max}\{ \sqrt {k_{max}}, \langle k^2\rangle/\langle k\rangle \} \ ,
\end{equation}
where the $o(1)$ term tends to zero as the limiting values
${\rm max}\{ \sqrt {k_{max}}, \langle k^2\rangle/\langle k\rangle \}$
diverge to infinity. 

For the classical, random ER graph case, with finite
connection probability: $p = \langle k \rangle  /N$ we have
$k_{max} = \ln N / \ln\ln N$, therefore 
\begin{equation}\label{L1scal}
\Lambda_1(N) = \big(1 + o(1) \big) {\rm max}\{ \sqrt {\ln N / \ln\ln N}, Np \} \ .
\end{equation}
Although in the $N\to\infty$ limit the first term dominates the maximum and 
predicts a divergence
\begin{equation}\label{L11scal}
\Lambda_1(N) = \sqrt {\ln N / \ln\ln N}  \ ,
\end{equation}
this function grows so slowly, that even for extremely large sizes:
$\Lambda_1(10^9) < 2.7$ and practically one observes a constant value. 
Thus, for $\langle k \rangle \ge 3$ and for $N < 2\times 10^5$ graphs
considered here the largest eigenvalue seems to tend to the finite value
\begin{equation} \label{L1k}
\Lambda_1(N) = \big(1 + o(1) \big) \langle k \rangle \ ,
\end{equation}
in agreement with the HMF theoretical value \cite{pv01a}:
\begin{equation}
\lambda_c^{HMF} = \langle k\rangle / \langle k^2\rangle 
\end{equation}
and with recent simulation results of Ref.~\cite{BCR13}.

For the random, unweighted SF networks, with power-law degree distribution 
$P(k)\propto k^{-3}$ the largest eigenvalue diverges and follows
the finite size scaling law
\begin{equation}\label{Lscal}
\Lambda_1(N) \propto N^{1/4} 
\end{equation}
deduced in \cite{HGVG13}.
Generally, in the numerical analysis of the QMF results least-squares fitting 
with the simple power-law form
\begin{equation} \label{lamscal}
1/\Lambda_1 = \lambda_c + Y (1/N)^{c}
\end{equation}
was applied, but in the ER case the logarithmic convergence 
of the largest eigenvalue (\ref{L11scal}) has been tested. 

\section{SIS model simulations}

Simulations of the SIS model were performed in such a way that in a given
time step either a deactivation at site $i$, with probability: $1/(1+\lambda)$ 
or activation of all neighboring inactive sites with probabilities:
$w_{ij}\lambda/(1+\lambda)/N_i$ were attempted.
Here $N_i$ is the number of neighboring, inactive sites, which was computed 
when the node $i$ was selected randomly. These reaction steps were iterated
$N_a$ times, where $N_a$ is the number of active sites at time $t$.
Following this system update, which selects each node once in average
the time was incremented by $\Delta t=1$ Monte Carlo step (MSc). 
Throughout the paper the time is measured by MCs.
The density of active sites $\rho(t)$ was calculated and stored at 
exponentially growing time steps: $t_i = 1 + 1.08^i$.
The system was initialized from a fully active state and the graph updates 
were repeated until $t < t_{max}$ or in case of extinction of activity.
The maximum simulation time depends on the system size, ranging from
$t_{max}=10^7$ for $N=10^5$ to $t_{max}=10^5$ for $N=10^6$.
To obtain good statistics the simulations have been repeated for $10^2-10^4$
independent graph realizations and  $\rho(t)$ averaged over them.

To explore in more detail the decay of the density functions, I have 
computed effective decay exponents of the power-laws 
$\rho(t) \propto t^{-\alpha}$, defined as the local slopes
\begin{equation}  \label{aeff}
  \alpha_{\rm eff}(t) = - \frac {\ln[\rho(t)/\rho(t')]} 
  {\ln(t/t^{\prime})} \ ,
\end{equation}
where $t$ and $t^{\prime}$ was chosen in such a way that the
discrete approximate of the derivative is sufficiently smooth.

\section{The SIS model on ER graphs} \label{sec:sis-er}

In the Erd\H os-R\'enyi graph a giant connected component emerges for
$\langle k\rangle \ge 1$ \cite{MR95}. Above this phase transition point
arbitrarily large connected sub-graph may exists and $d=\infty$. 
It has been conjectured \cite{PhysRevLett.105.128701,Juhasz:2011fk} 
that in this case the epidemic spreading is too fast to let the formation 
of active RRs of size $l$ with the lifetime: $\tau\propto\exp(l)$, 
hence GP cannot occur.
Contrary, for $\langle k\rangle < 1$ the topological dimension is zero
in the ER graph.
It was also hypothesized \cite{PhysRevLett.105.128701,Juhasz:2011fk} and 
shown by simulations of the CP \cite{ICNPproc}, that in the fragmented 
phase strong rare-region effects and GP dynamics can be observed.

\begin{figure}[ht]
\includegraphics[height=6cm]{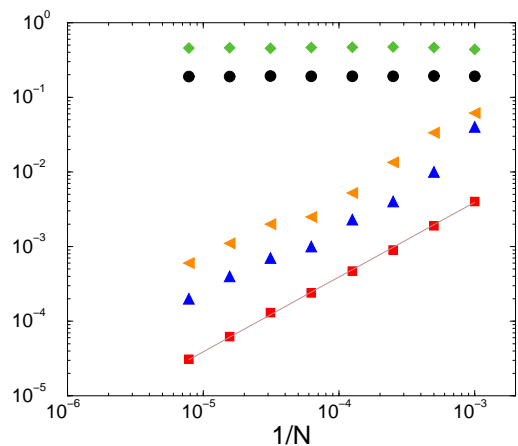}
\caption{\label{er4} (Color online) Finite size scaling of QMF results 
  on the ER model with $\langle k\rangle = 4$ for $N=1000$, $2000$, $4000$, ..., $128000$. 
  Bullets, $\lambda_c$; boxes, IPR; up-triangles, $a_1$; down-triangles, 
  $a_2$; right-triangles, $a_3$.
  Line: least-squares fitting with the form $\sim 1/N$.}
\end{figure}

Now I investigate this hypothesis by simulations and with the help of 
QMF method in case of the SIS model. Application of the QMF leads to 
the following results. 
In the percolating phase with $\langle k\rangle = 4$ the $IPR(N)$ value
decays to zero as $\sim 1/N$, indicating the disappearance of activity 
clustering in the steady state as shown on Fig.~\ref{er4}.
On the other hand $a_1$ is roughly constant, while $a_2$ and $a_3$ 
extrapolate to zero, suggesting a clean mean-field critical transition, 
characterized by $\beta=1$ leading order parameter exponent in agreement 
with our expectations. 
In the $N\to\infty$ limit extrapolation with the form (\ref{Lscal}) results
in $\Lambda_1 = 5.23(3)$, which is larger than what a simulation should
show for reasonable graph sizes ($N<10^9$): $\Lambda_1 = 4$ 
(see Eq.~(\ref{L1k})).
The critical point estimate of QMF is: $\lambda_c = 1/\Lambda_1 = 0.191(1)$.
Probably the application of pair, or higher level QMF, taking into account 
longer correlations of the order parameter would increase this value as 
in \cite{MF13} for random regular networks. These results 
are summarized in the first line of Table~\ref{tabla}.

\begin{figure}[t]
\includegraphics[height=6cm]{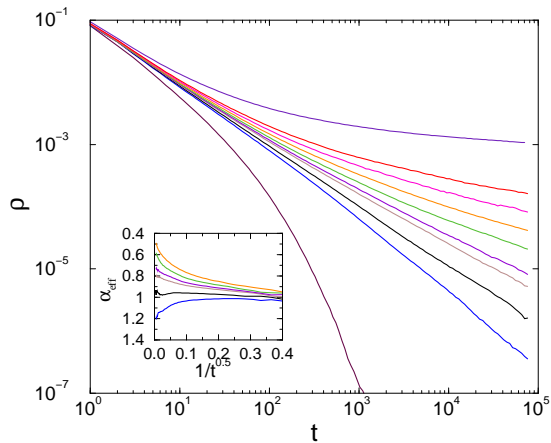}
\caption{\label{ersis1} (Color online) Density decay as a function of time 
  for the SIS on ER graph with
  $\langle k\rangle = 1$. Network size $N=10^6$. Different curves correspond to
  $\lambda=$ 1.05, 1.09, 1.095, 1.1, 1.02, 1.05, 1.095, 1.115, 1.12, 1.15 
   (from bottom to top curves). Inset: effective exponents of the same data 
   near the phase transition point.}
\end{figure}

Extensive simulations for the more interesting, limiting case 
$\langle k\rangle = 1$ have been performed for graphs with 
$N=10^6$ nodes. As Figure~\ref{ersis1} shows a
mean-field type of phase transition appears at $\lambda_c =1.094(1)$, 
which is close to the the HMF value $\lambda_c^{HMF}=1$.
 
\begin{figure}[ht]
\includegraphics[height=6cm]{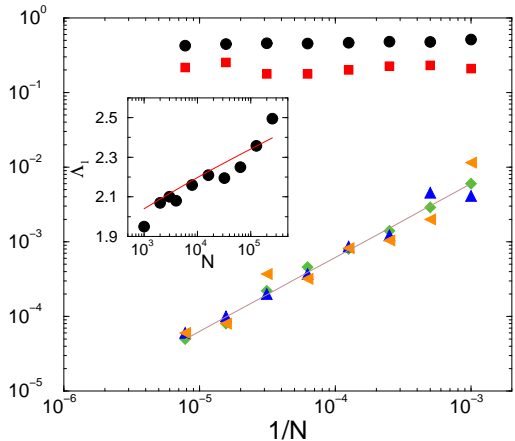}
\caption{\label{er1} (Color online)
  Finite size scaling of QMF results for ER graph with $\langle k\rangle = 0.3$
  $N=1000$, $2000$, $4000$, ... $128000$. Bullets, $\lambda_c$; boxes, IPR; 
  up-triangles, $a_1$; down-triangles, $a_2$; right-triangles, $a_3$.
  Line: least-squares fitting with a form $1/N$.
  Inset: $\Lambda_1(N)$ (bullets) and the form (\ref{L11scal}) (line).
}
\end{figure}

In the fragmented phase, for $\langle k\rangle = 0.3$ the IPR remains constant
and tends to $0.22(2)$ (see Fig.~\ref{er1}) and the coefficients of the
order parameter $\rho(\lambda)$ vanish as $a_i \propto 1/N$. 
The epidemic threshold estimate $\lambda_c = 1/\Lambda_1$ decreases very slowly
with $N$. The inset of Fig.~\ref{er1} shows the slowly increasing $\Lambda_1$
approximated with the form (\ref{L11scal}). 
However, the density fluctuations of $\rho$ drive finite clusters,
thus the whole system into the absorbing state in the fragmented phase,
that can't be described by the QMF. Since a real GP singularity causes 
dynamical behavior even in the active phase the clustering behavior 
obtained by the SD decomposition predicts the existence of strong 
RR effects correctly. This underlines the capability of QMF to treat
the effects of topological disorder well. 
The numerical results are summarized in the row 'ER-03' of Table~\ref{tabla}.

\begin{figure}[t]
\includegraphics[height=6cm]{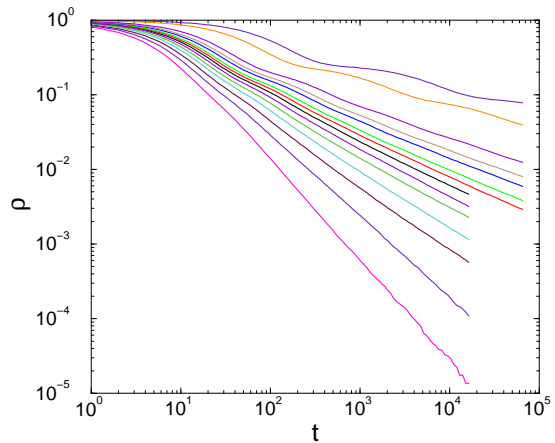}
\caption{\label{ersis03} (Color online)
  Density decay as a function of time in the SIS defined on ER graph with
  $\langle k\rangle = 0.3$. Network size $N=10^6$. Different curves correspond to
  $\lambda=$ 4, 5, 6, 7, 8, 9, 10, 11, 12, 14, 16, 20, 50, 100 
  (from bottom to top curves).}
\end{figure}

The simulations confirm the existence of GP by showing an extended 
region of $\lambda$ dependent power-laws of the density decay. 
As one can see of Fig.~\ref{ersis03} for $\langle k\rangle=0.3$ 
the slopes of the curves on the log.-log. plot vary in wide range from 
$\simeq 1.34(1)$ at $\lambda=4$ to $\simeq 0.43(2)$ at $\lambda=20$. 
For higher $\lambda$-s the decay curves show oscillations superimposed 
on the power-law. 
This can be explained by the gradual deactivation of RR-s in the fragments.

Finally, I have also studied the steady state behavior by QMF on 
a weighted ER model introduced in \cite{b13}.
This model was motivated by the face-to-face experiments analyzed in \cite{F2F} 
with infection rates $\lambda w_{ij}$, proportional to the intensity of 
contacts, such that the quenched weights generate a 
$P(\omega) = 1/\omega$ probability distribution function. 
In this model the both the simulations, HMF and percolation approaches 
\cite{b13} suggests RR effects in the percolating phase of ER with 
$\langle k\rangle = 4$. Weight factors are displaced on the graph edges
\begin{equation} \label{expw}
w_{ij} = \exp{(-a r)} \ \ ,
\end{equation} 
where $r\in (0,1)$ is a uniformly distributed random variable and $a$ controls the 
strength of the disorder $w_{ij} \in (1/e^{a}, 1)$ as in Ref.~\cite{b13}.

\begin{figure}[ht]
\includegraphics[height=6cm]{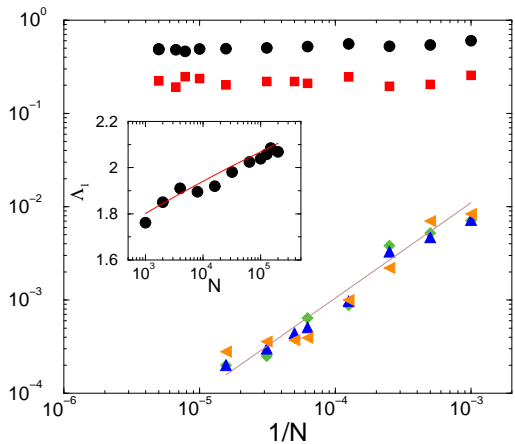}
\caption{\label{er4-6} (Color online)
  Finite size scaling of QMF results for the weighted ER graphs 
  for $\langle k\rangle = 4$, in the range of sizes $N=1000$, $2000$ ...$200000$.
  Bullets, $\lambda_c$; boxes, IPR; up-triangles, $a_1$; down-triangles, 
  $a_2$; right-triangles, $a_3$. Line: least-squares fitting with a $1/N$.
  Inset: $\Lambda_1(N)$ (bullets) and the form (\ref{L11scal}) (line). 
}
\end{figure}

The QMF analysis for $a=6$ shows that this kind disorder suppresses the infection 
rates very efficiently and one arrives to very similar results as for the 
fragmented case (see Fig.~\ref{er4-6}).
The $IPR(N)$ values are finite and constant the coefficients $a_i$ decay as 
$\sim 1/N$ and $\Lambda_1(N)$ follows the theoretical form (\ref{L11scal})
(see inset of  Fig.~\ref{er4-6}).
The $N\to\infty$ extrapolated values are summarized in the row
'ER-4-6' of Table~\ref{tabla}. This means that due to the lower cutoff
value of the weights the network is quasi fragmented and the RR effects 
reported in \cite{b13} are supported by the QMF method.

On the other hand, if the disorder is weak, characterized by $a=1$ a 
similar QMF analysis leads to the values shown in the row 'ER-4-E1' 
of Table~\ref{tabla}.
In this case $\Lambda_1(N)$ remains finite, while the $IPR(N)$ 
converges to zero, meaning the lack of RR effects in agreement
with the phase diagram shown in \cite{b13}.

\section{The SIS model on aging BA graphs} \label{sec:sis-ba}

In this section I show the application of the QMF method to generalized 
BA tree networks (BAT) in which the spreading behavior is weakened by 
preferential depletion of the links \cite{SAH11}.
In Ref.~\cite{wbacikk} SIS models on top of BA graphs have been 
studied by QMF and extensive dynamical simulations. 
For these infinite topological dimensional networks no strong 
rare-region effects have been found, except when dissortative 
weighting scheme was applied, which suppresses the hub-hub connections. 
This slows down the fast epidemic spreading in case of tree 
BA topologies.

Now I consider the SIS model on generalizations of 
BA type of networks \cite{Barabasi:1999}.
The choice of this model is motivated by the fact that it allows to
construct tree structures in a very simple way, in contrast with
other standard network generation models, e.g \cite{ucmmodel}. 
BA is a growing network model in which, at each time step $s$, a new
vertex (labeled by $s$) with $m$ edges is added to the network 
and connected to an existing vertex $s'$ of degree $k_{s'}$ with probability 
$\Pi_{s \rightarrow s'} = k_{s'} /\sum_{s''<s} k_{s''}$. This process is
iterated until reaching the desired network size $N$. The resulting
network has a SF degree distribution $P(k) \simeq 2 m^2 k^{-3}$;
additionally, fixing $m=1$ leads to a strictly tree (loop-less)
topology.

\begin{figure}[t]
\includegraphics[height=6cm]{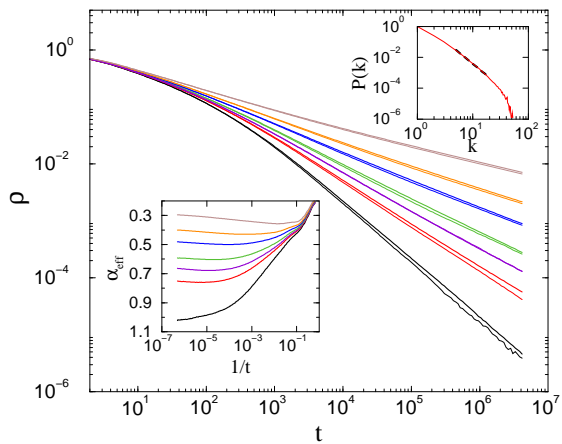}
\caption{\label{BAsisg2m} (Color online)
  Density decay as a function of time for the SIS on BA graph with
  network sizes: $N=10^5$ (thin lines), $N=10^6$ (thick lines). 
  Different curves correspond to $\lambda=$ 2.4, 2.45, 2.47, 2.5, 2.55, 2.6, 2.7 
  (from bottom to top curves). 
  Left inset: Local slopes of the same data. Right inset: Degree distribution
  of the aging BA graph for $N=4\times 10^6$ nodes.
}
\end{figure}

Following the generation of BAT an aging procedure was applied 
on the network by gradually removing a fraction of randomly selected
links connecting nodes $i$ and $j$ with the probability
\begin{equation}
q_{i,j} = (k_i k_j) / \sum_{i=1}^N k_i \ ,
\end{equation}
The procedure was repeated until $ \sum_i k_{i=1}^N \ge 1.6 N$, i.e. $20\%$
of the original links of the graph were removed. As the consequence the
original scale-free distribution $P(k) \propto k^{-3}$ gets an 
exponential cutoff as shown in the right inset of Fig.~\ref{BAsisg2m}. 
This is quite similar to what one observes in most empirical degree 
distributions. The topological dimension of the graph becomes finite.

\begin{figure}[ht]
\includegraphics[height=6cm]{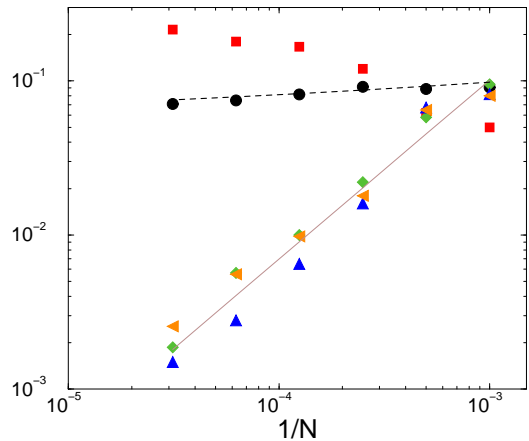}
\caption{\label{bag2m} (Color online)
  Finite size scaling of QMF SD results of the SIS on generalized BAT
  with $N=1000$, $2000$, $4000$, $8000$, $16000$, $32000$
  nodes. Bullets, $\lambda_c$; boxes, IPR; up-triangles, $a_1$; 
  down-triangles, $a_2$; right-triangles, $a_3$.
  Line: least-squares fitting with the form $\sim 1/N$.
}
\end{figure}

The SD analysis of this model (GBA2) gives IPR results with strong clustering 
in the $N\to\infty$ limit, thus it suggests rare-region effects 
(see Fig.~\ref{bag2m}).
While the $IPR$ converges to $0.18(5)$ the largest eigenvalue
increases and the coefficients $a_i$ decrease as $\sim 1/N$. 
A power-law fit for the critical point estimate results in 
$\lambda_c = 1/\Lambda_1 = 0.001 + 0.17 (1/N)^{0.18}$, which is a 
similar to the behavior found in case of SIS on BA graphs \cite{wbacikk}. 
The finite size scaling exponent $c$ is close to the value predicted 
by analytical considerations see Eq.~((\ref{Lscal})).

Simulations show clear $\lambda$-dependent, power-law density decays 
(see Fig.~\ref{BAsisg2m}). The effective decay exponents saturate to
constant values in the large time limit (left inset of  Fig.~\ref{BAsisg2m})
albeit some (log.) correction is possible. Furthermore, the results 
do not seem to depend on the size of the networks used. These are
well known indications of Griffiths Phases.
 
\begin{table}
\caption{Spectral QMF analysis results for SIS in different networks\label{tabla}}
\begin{center}
\begin{tabular}{|l|r|r|r|r|}
\hline
Network & $1/\Lambda_1$ & $c$		&     IPR  	 &	$b$		\\
\hline
ER-4     & $0.19(1)$	& $1.1(1)$  	& $0.00003(5)$   &	$1.1(1)$ \\
ER-03    & $0.01(1)$    &    -          & $0.22(2)$     &       $0.9(2)$  \\
ER-4-E6  & $0.01(1)$	&    -  	& $0.22(3)$	 &	$0.9(3)$ \\
ER-4-E1  & $0.30(3)$	& $0.34(1)$  	& $0.01(1)$ 	&	$0.93(9)$ \\
GBA2	& $0.001(2)$	& $0.18(5)$  	& $0.28(5)$	& $0.5(1)$	\\
\hline
\end{tabular}
\end{center}
\end{table}

\section{Conclusions}

In conclusion the usability of the QMF method for detecting rare-region
effects has been demonstrated on different complex networks. 
Comparison with dynamical simulations has confirmed that the clustering 
behavior, in the steady-state is related to the occurrence of slow behavior 
and Griffiths Phases. 
Finite size scaling and asymptotic values of the largest eigenvalue of 
the adjacency matrix agree with the theoretical predictions for the 
SIS model defined on top of ER and BA networks.
In particular the slow, logarithmic decrease of the threshold value for
random networks is studied. In the fragmented phase QMF results in 
inhomogeneous, clustered principal eigenvector. 
However, it can't describe well the inactive phase, because it neglects 
dynamical fluctuations, which in a real system drive all finite 
clusters to extinction in finite times. Still, it predicts correctly the 
relevancy of topological disorder in agreement with the GP found 
by the simulations.
In the percolating phase of ER graph the QMF method results in 
the irrelevancy of network heterogeneity on the SIS scaling, 
providing the expected the mean-field critical behavior. This
has been confirmed by dynamical simulations at $\langle k\rangle =1$.
  
The methodology has been applied to two more interesting cases, where
GP behavior could be expected. In case of a quenched weight 
scheme, capable of describing phenomenologically observed face-to-face 
inter-event distribution QMF supports recent numerical simulation results.
However, it is more likely than the dynamical power-law behavior is
restricted to finite times, because in an infinite dimensional graph
arbitrarily large (dimensional) subspaces may survive an extinction
process and provide finite contribution to the density.
For weak disorders it predicts the lack of GP on the percolating ER model.
It has also been shown that QMF describes well the GP behavior of SIS on 
a Barab\'asi-Albert type model with aging nodes, i.e. when links are diluted 
by a preferential detachment rule. This is confirmed by dynamical 
simulations, exhibiting non-universal power-laws, with negligible size dependencies.

Finite size scaling of the QMF results also predict the disappearance of the 
largest three amplitudes ($a_i$) of the order parameter as $\sim 1/N$,
except for the non-clustering ER cases, when the leading order one 
($a_1$) remains constant, meaning a linear mean-field exponent with
negligible corrections.
Application of the present method to real networks, where large-scale
modularity is an ubiquitous property \cite{N11} would reveal interesting
consequences on the dynamics of processes evolving on them.

\section*{Acknowledgments}

I thank R. Juh\'asz for useful discussions.
Support from the Hungarian research fund OTKA (Grant No. T77629) 
and the European Social Fund through project FuturICT.hu (grant no.:
TAMOP-4.2.2.C-11/1/KONV-2012-0013) is acknowledged.

\end{document}